\documentclass[prl,aps,showpacs,floatfix,nofootinbib,letterpaper,twocolumn,reprint]{revtex4-1}
\usepackage{epsfig} 
\usepackage{amssymb} 
\usepackage{amsmath} 
\usepackage{amsfonts} 
\usepackage{color, graphicx} 
\usepackage{bm}

\def\lb{\langle}
\def\rb{\rangle} 
 
\def\be{\begin{equation}} 
\def\ee{\end{equation}} 

\def\Tr{{\rm Tr}}

\def\Im{{\rm Im\,}}

\newcommand{\avg}[1]{\lb #1 \rb}

\begin{document}

\title{Low-energy enhancement in the magnetic dipole $\gamma$-ray strength functions of heavy nuclei}
\author{P. Fanto}\email{paul.fanto@yale.edu}
\author{Y. Alhassid}\email{yoram.alhassid@yale.edu}
\affiliation{Center for Theoretical Physics, Sloane Physics Laboratory, Yale University, New Haven, Connecticut 06511, USA}
\date{\today}

\begin{abstract}
A low-energy enhancement (LEE), observed experimentally in the $\gamma$-ray strength function ($\gamma$SF) describing the decay of compound nuclei, would have profound effects on $r$-process nucleosynthesis if it persists in heavy neutron-rich nuclei.  The LEE was shown to be a feature of the magnetic dipole ($M1)$ strength function in configuration-interaction shell-model calculations in medium-mass nuclei. However, its existence in heavy nuclei and its evolution with neutron number remain open questions. Here, using a combination of many-body methods, we find the LEE in the $M1$ $\gamma$SFs of heavy samarium nuclei.   In particular, we use the static-path plus random-phase approximation (SPA+RPA), which includes static and small-amplitude quantal fluctuations beyond the mean field.  Using the SPA+RPA strength as a prior, we apply the maximum-entropy method (MEM) to obtain finite-temperature $M1$ $\gamma$SFs from exact imaginary-time response functions calculated with the shell model Monte Carlo (SMMC) method.  We find that the slope of the LEE  in  samarium isotopes is roughly independent of the average initial energy over a wide range below the neutron separation energy.  As the neutron number increases, strength transfers to a low-energy excitation, which we interpret as the scissors mode built on top of excited states.  
\end{abstract}

\maketitle

{\it Introduction} -- The $\gamma$-ray strength function ($\gamma$SF) \cite{barth1973} is an important input to the Hauser-Feshbach theory of  compound-nucleus reactions~\cite{koning2012} and has a significant effect on $r$-process nucleosynthesis \cite{mumpower2016}.   In particular, the magnetic dipole ($M1$) $\gamma$SF in heavy nuclei exhibits interesting phenomenology, characterized by a spin-flip resonance and a ``scissors" mode 
~\cite{bohle1984,heyde2010}.  
The inclusion of the latter in Hauser-Feshbach calculations improves predictions of neutron radiative capture rates~\cite{mumpower2017}. 

In recent years, a low-energy enhancement (LEE) -- a so-called ``upbend'' structure at low $\gamma$-ray energies -- has been observed experimentally in the $\gamma$SFs for decay in several mid-mass nuclei~\cite{voinov2004,larsen2013,larsen2018} and a few rare-earth nuclei~\cite{simon2016,naqvi2019}.  If it persists in heavy neutron-rich nuclei, the LEE is likely to have profound effects on $r$-process nucleosynthesis since it would enhance significantly the radiative neutron capture cross sections of nuclei near the neutron drip line~\cite{larsen2010}. Experiments indicate that the LEE is of dipole nature~\cite{jones2018}, and configuration-interaction (CI) shell-model studies of medium-mass nuclei have attributed the LEE to the $M1$ $\gamma$SF~\cite{schwengner2013, brown2014, karampagia2017, schwengner2017, sieja2017, mitdbo2018}.  However, conventional CI shell-model diagonalization methods are limited to light and mid-mass nuclei due to the combinatorial increase of the many-particle model space with the numbers of valence nucleons and/or single-particle orbitals.  Consequently, there has been no theoretical confirmation of the persistence of the LEE in heavy nuclei. 

While most theoretical studies have focused on the strength function built on the ground state, finite-temperature methods enable the study of $\gamma$SFs built on excited states.  The finite-temperature quasiparticle random-phase approximation (QRPA) has been applied to calculate finite-temperature $\gamma$SFs~\cite{litvinova2013,litvinova2018,wibowo2019,yuksel2020}, and the zero-temperature QRPA with empirical corrections has also been applied to $\gamma$SF calculations~\cite{martini2016,goriely2018}.  However, the QRPA has limitations in that it only includes small-amplitude quantal fluctuations around the mean-field configuration, and its finite-temperature version has been mostly limited to spherical nuclei.

The shell-model Monte Carlo (SMMC) method ~\cite{johnson1992, alhassid1994,koonin1997, alhassid2008} enables exact (to within statistical errors) calculations of finite-temperature observables within the CI shell-model framework; see Ref.~\cite{alhassid_rev} for a recent review.  However, the SMMC cannot calculate the finite-temperature $\gamma$SF directly, but only its Laplace transform, the imaginary-time response function.  

Here we combine the static-path plus random-phase approximation (SPA+RPA)~\cite{puddu1991, attias1997, rossignoli1998_strength} with the exact SMMC method to calculate finite-temperature $M1$ $\gamma$SFs of heavy nuclei.  The SPA+RPA includes large-amplitude static fluctuations~\cite{lauritzen1988} and small-amplitude time-dependent quantal fluctuations beyond the mean field and was recently successfully applied to study nuclear state densities of heavy nuclei~\cite{fanto2021}.  To improve on the SPA+RPA strength functions, we apply the maximum-entropy method (MEM)~\cite{gubernatis1991, jarrell1996,gubernatis_book} to obtain finite-temperature $\gamma$SFs from the SMMC imaginary-time response functions, using the SPA+RPA strength as a prior.

We apply this method to calculate finite-temperature $M1$ $\gamma$SFs in a chain of heavy even-mass samarium isotopes $^{148-154}$Sm.  Furthermore, we extract average reduced transition probabilities ($B(M1)$ values) from these finite-temperature strength functions. 
In the average $B(M1)$ values for decay, we find an enhancement at low $\gamma$-ray energies that is present over a broad range of temperatures (i.e., average initial excitation energies), which we interpret as the LEE. 

 In the  finite-temperature $M1$ $\gamma$SFs, we also observe a transfer of strength from the low-energy peak to a somewhat higher-energy excitation as the neutron number increases and the isotopes become more deformed.  We interpret this higher-lying excitation as the ``scissors'' mode~\cite{bohle1984,heyde2010} built on excited states~\cite{krticka2011}.

{\it SPA+RPA $\gamma$SF} -- The $\gamma$SF at temperature $T$ for an electromagnetic  transition operator, which is a spherical tensor $\mathcal{O}_\lambda$ of rank $\lambda$, is defined by
\begin{eqnarray}\label{strength_cism}
S_{\mathcal{O}_\lambda}(T; \omega) = \sum_{\substack{\alpha_i J_i \\ \alpha_f J_f}} \frac{e^{-\beta E_{\alpha_i J_i}}}{Z} |(\alpha_f J_f || \hat{\mathcal{O}}_\lambda|| \alpha_i J_i  )|^2 \nonumber \\ \times \delta(\omega - E_{\alpha_f J_f}+ E_{\alpha_i J_i })\,,
\end{eqnarray}
where $(\alpha J)$ label CI shell-model eigenstates with energy and spin $(E_{\alpha J}, J)$, and $Z = \sum_{\alpha J} (2J+1) e^{-\beta E_{\alpha J}}$ is the canonical partition function.  
Given a shell-model Hamiltonian with the two-body interaction expressed in the separable form $\hat H_2 = -(1/2)\sum_\alpha v_\alpha \hat Q_\alpha^2$, we apply the adiabatic approximation of Ref.~\cite{rossignoli1998_strength} to evaluate Eq.~(\ref{strength_cism}) in the SPA+RPA
\be\label{spa_rpa_gsf}
S_{\mathcal{O}_\lambda}(T;\omega) \approx \frac{\int d\sigma M(\sigma) Z_\eta(\sigma) C_\eta(\sigma) S_{\mathcal{O}_\lambda, \eta}(T,\sigma;\omega)}{\int d\sigma M(\sigma) Z_\eta(\sigma) C_\eta(\sigma)}\,,
\ee
where $\sigma$ are static auxiliary fields and $M(\sigma)$ is a measure function~\cite{fanto2021}. $Z_\eta(\sigma) = \Tr\left[ \hat P_\eta e^{-\beta \left(\hat h_\sigma - \sum_{\lambda=p,n} \mu_\lambda \hat N_\lambda\right)}\right]$ is the number-parity projected one-body partition function, where $\hat h_\sigma = \hat H_1 - \sum_\alpha v_\alpha \sigma_\alpha \hat Q_\alpha$ is a one-body Hamiltonian, and $P_\eta=(1+\eta e^{i\pi \hat N})/2$ is the number-parity projection with $\eta = +1(-1)$ for even (odd) number parity.  $C_\eta(\sigma)$ is the RPA correction factor that accounts for the Gaussian integral over small amplitude time-dependent auxiliary-field fluctuations~\cite{puddu1991, attias1997, rossignoli1998, nesterov2013, fanto2021}. This correction factor is given by (see Eq.~(8) of Ref.~\cite{fanto2021})
\be\label{rpa_corr}
C_\eta(\sigma) = \frac{\prod_{k > l} \frac{1}{\tilde E_k - \tilde E_l} \sinh\left(\beta(\tilde E_k - \tilde E_l)/2\right)}{ \prod_{\nu > 0} \frac{1}{\Omega_\nu} \sinh\left(\beta \Omega_\nu/2\right)}\,,
\ee
where $\tilde E_k$ are the generalized quasiparticle energies of $\hat h_\sigma$ and $\pm \Omega_\nu$ are the eigenvalues of the $\sigma$-dependent RPA matrix
\be\label{rpa_matrix}
\mathcal{M}_{kl,k^\prime l^\prime}^\eta = (\tilde E_k - \tilde E_l)\delta_{kk^\prime} \delta_{ll^\prime} - \frac{1}{2}(\tilde f^\eta_l - \tilde f^\eta_k)\sum_{\alpha} \mathcal{Q}_{\alpha,kl} \mathcal{Q}_{\alpha,l^\prime k^\prime}\,.
\ee
Here $\tilde f^\eta_k$ are the number-parity-projected generalized thermal quasiparticle occupation numbers; see Ref.~\cite{fanto2021} for further details.

The $\sigma$-dependent strength function $S_{\mathcal{O}_\lambda, \eta}(T,\sigma;\omega)$ in Eq.~(\ref{spa_rpa_gsf}) is given by~\cite{rossignoli1998_strength}
\be\label{sigma_gsf}
\begin{split}
S_{\mathcal{O}_\lambda,\eta}(T,\sigma;\omega) & = S_{\mathcal{O}_\lambda,\eta}^{(0)}(T,\sigma)\delta(\omega) \\
& - \lim_{\epsilon \to 0^+} \frac{\Im \Pi_{\mathcal{O}_\lambda,\eta}(T,\sigma; \omega + i\epsilon)}{\pi(1 - e^{-\beta \omega})} \,,
\end{split}
\ee
where
\begin{eqnarray}\label{dcf}
\Pi_{\mathcal{O}_\lambda,\eta}(T,\sigma; \omega + i\epsilon)  &= \sum_\mu \sum_{kl, k^\prime l^\prime}\sum_{\nu} \frac{1}{2}\mathcal{O}_{\lambda\mu,kl}^* \mathcal{M}_{kl,\nu}^\eta  \mathcal{M}^{\eta,-1}_{\nu, k^\prime l^\prime} \nonumber \\
&\times (\tilde f^\eta_{l^\prime} - \tilde f^\eta_{k^\prime}) \mathcal{O}_{\lambda \mu,k^\prime l^\prime} \frac{1}{\omega - \Omega_\nu + i\epsilon}\,.
\end{eqnarray}
In Eq.~(\ref{dcf}), $\nu$ ranges over the eigenbasis of the RPA matrix $\mathcal{M}$ and $\mu$ denotes the components of 
$\mathcal{O}_\lambda$.  The second term on the right-hand side of Eq.~(\ref{sigma_gsf}) corresponds to the finite-temperature QRPA strength function around the mean-field configuration defined by the static auxiliary fields $\sigma$~\cite{sommermann1983, ring1984}.  The first term on the right-hand side accounts for the strength in the $\omega \to 0$ limit.  The RPA correlations are challenging to take into account for this term, and we use the SPA result~\cite{rossignoli1998_strength}
\be\label{S0}
S_{\mathcal{O}_\lambda,\eta}^{(0)}(T,\sigma) = \frac{1}{2}\sum_\mu \sum_{\substack{kl,k^\prime l^\prime\\ E_k = E_l}} \mathcal{O}_{\lambda\mu,kl}^* \mathcal{O}_{\lambda \mu,k^\prime l^\prime} \langle a_k^\dagger a_l a_{k^\prime}^\dagger a_{l^\prime}\rangle_{\sigma,\eta}\,,
\ee
where $a_k, a_k^\dagger$ annihilate and create quasiparticles in the eigenbasis of $\hat h_\sigma$. 

We express Eq.~(\ref{spa_rpa_gsf}) in terms of the weight function $\mathcal{W}_\eta(\sigma) = M(\sigma) Z_\eta(\sigma)$
\be\label{gsf_mc}
S_{\mathcal{O}_\lambda}(T;\omega) = \frac{\int d\sigma \mathcal{W}_\eta(\sigma) C_\eta(\sigma) S_{\mathcal{O}_\lambda, \eta}(T,\sigma;\omega)}{\int d\sigma \mathcal{W}_\eta(\sigma) C_\eta(\sigma)}\,.
\ee
Following Ref.~\cite{fanto2021}, we apply the Metropolis-Hastings algorithm to draw uncorrelated sample configurations $\sigma_k$ from the weight function $\mathcal{W}_\eta$ and evaluate Eq.~(\ref{gsf_mc}) by taking an average over these sample configurations,

We also calculate the finite-temperature $\gamma$SF in the static-path approximation (SPA), which neglects the quantal fluctuations entirely.  In this case, the RPA correction factor $C_\eta =1$, and the $\sigma$-dependent $\gamma$SF in Eq.~(\ref{sigma_gsf}) is given by its mean-field expression. 

\begin{figure*}[]
\includegraphics[width=\textwidth]{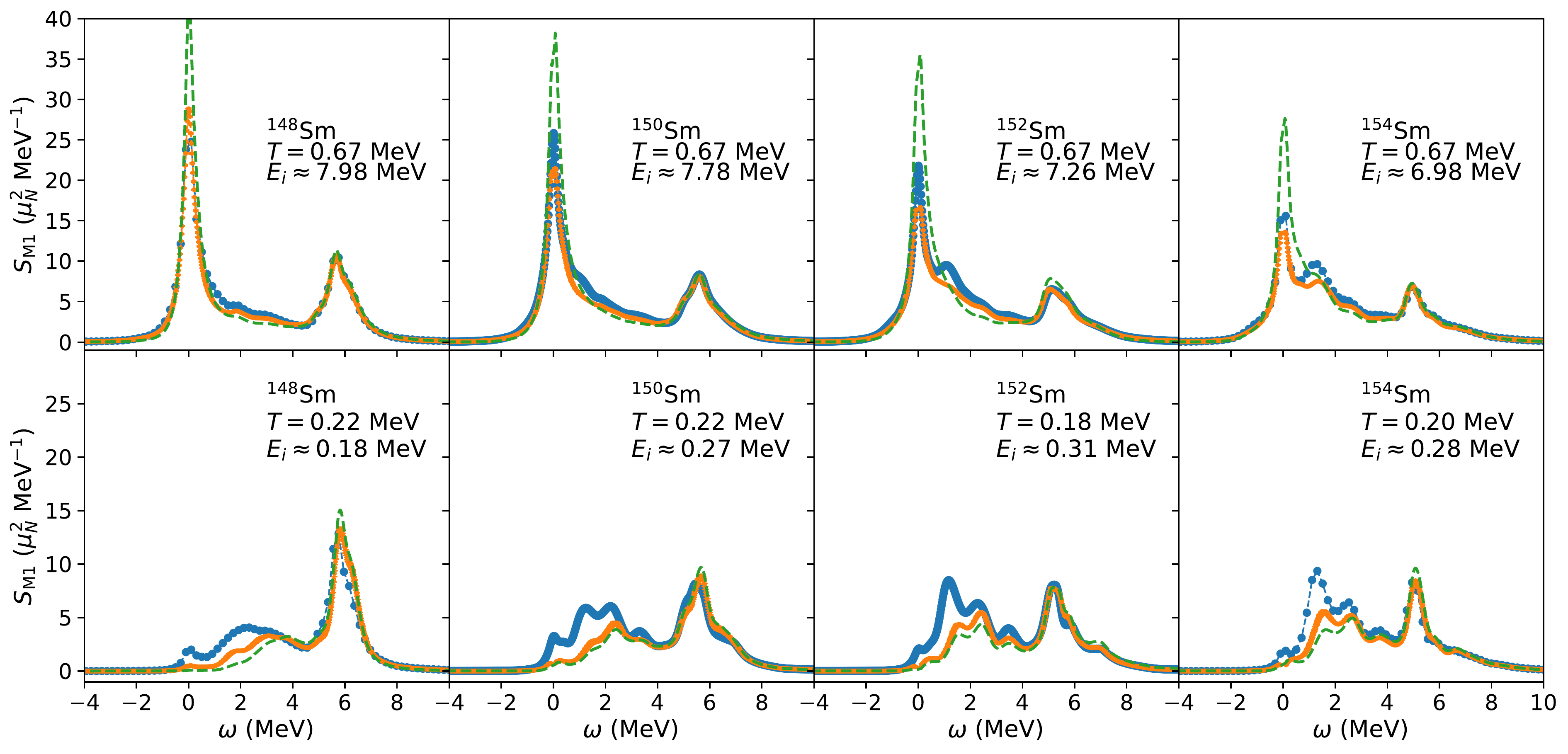}
\caption{\label{gsf-figure} The finite-temperature $M1$ $\gamma$SF as a function the transition energy $\omega$ for the even-mass samarium isotopes $^{148-154}$Sm.  The MEM results (green dashed lines) are compared with the SPA+RPA results (orange dots) and SPA results (blue circles).  The top row shows temperatures near the neutron separation energies in each of the isotopes, while the bottom row shows low temperatures close to the ground state.}
\end{figure*}

{\it Maximum-entropy method} -- We improve our SPA+RPA calculations of the finite-temperature $\gamma$SF by combining these results with exact SMMC calculations of the imaginary-time response function  $R_{\mathcal{O}_\lambda}(T; \tau)$ through the maximum-entropy method (MEM)~\cite{jarrell1996}.   $R_{\mathcal{O}_\lambda}(T; \tau)$ is the Laplace transform of the finite-temperature $\gamma$SF.  The $\gamma$SF satisfies $ S_{\mathcal{O}_\lambda}(T; -\omega)= e^{-\beta \omega} S_{\mathcal{O}_\lambda}(T; \omega)$ and the Laplace transform can be rewritten as
\be\label{response_integral}
R_{\mathcal{O}_\lambda}(T; \tau) = \int_0^\infty d\omega\, K(\tau,\omega) S_{\mathcal{O}_\lambda}(T; \omega)\,,
\ee
where $K(\tau,\omega)= e^{-\tau\omega} + e^{-(\beta-\tau) \omega}$ is a symmetrized kernel that decays exponentially with $\omega$. 
The MEM selects the $\gamma$SF that maximizes the objective function \cite{jarrell1996}
\be\label{Q_maxent}
\mathcal{Q}(S_{\mathcal{O}_\lambda}; \alpha) = \alpha \mathcal{S} - \frac{1}{2}\chi^2\,.
\ee
In Eq.~(\ref{Q_maxent}), the $\chi^2$ function is given by
\be\label{chi_sq}
\chi^2 = (\overline{R}_{\mathcal{O}_\lambda} - R_{\mathcal{O}_\lambda})^T \mathcal{C}^{-1} (\overline{R}_{\mathcal{O}_\lambda} - R_{\mathcal{O}_\lambda})\,,
\ee
where $\overline R$ is the SMMC response function and $\mathcal{C}$ is its covariance matrix.  $\mathcal{S}$ is the entropy function
\be\label{entropy}
\begin{split}
\mathcal{S} = & -\int d\omega\, \bigg( S_{\mathcal{O}_\lambda}(T; \omega) - S^{\rm prior}_{\mathcal{O}_\lambda}(T; \omega)\\
& - S_{\mathcal{O}_\lambda}(T; \omega)\ln \left[S_{\mathcal{O}_\lambda}(T; \omega)/S^{\rm prior}_{\mathcal{O}_\lambda}(T; \omega)\right]\bigg)\,,
\end{split}
\ee
where $S^{\rm prior}_{\mathcal{O}_\lambda}$ is a suitably chosen prior for the $\gamma$SF.  

In this work, we apply Bryan's method \cite{bryan1990}, in which the MEM strength function is given by
\be\label{strength_bryans}
S^{\rm MEM}_{\mathcal{O}_\lambda} = \int d\alpha \, S^\alpha_{\mathcal{O}_\lambda} \mathcal{P}(\alpha|\overline G_{\mathcal{O}_\lambda}, \mathcal{C}, S^{\rm prior}_{\mathcal{O}_\lambda})\,,
\ee
where $S^\alpha_{\mathcal{O}_\lambda}$ maximizes the objective function (\ref{Q_maxent}) for a given $\alpha$, and the probability function $\mathcal{P}(\alpha|\overline G_{\mathcal{O}_\lambda}, \mathcal{C}, S^{\rm prior}_{\mathcal{O}_\lambda})$ is given in Ref.~\cite{jarrell1996}.

In the SMMC method, the imaginary-time response function of $\mathcal{O}_\lambda$  is expressed as~\cite{koonin1997, alhassid_rev}
\be\label{response_smmc}
R_{\mathcal{O}_\lambda}(T; \tau) = \frac{\int D[\sigma] G_\sigma \Tr\, \hat U_\sigma  \avg{\mathcal{O}_\lambda(\tau) \cdot \mathcal{O}_\lambda}_\sigma }{\int D[\sigma] G_\sigma \Tr \,\hat U_\sigma}\,,
\ee
where $ \mathcal{O}_\lambda(\tau)=\hat U^{-1}_\sigma(\tau,0)  \mathcal{O}_\lambda \hat U_\sigma(\tau,0)$  with $\hat U_\sigma(\tau,0)$ being the propagator for a system of non-interacting nucleons moving in external time-dependent auxiliary fields $\sigma$, $G_\sigma$ is a Gaussian weight, and the expectation value $\langle ... \rangle_\sigma$ is taken with respect to the propagator $\hat U_\sigma \equiv \hat U_\sigma(\beta,0)$.  We sample auxiliary-field configurations according to the weight function $\mathcal{W}_\sigma^{\rm SMMC} = G_\sigma | \Tr \hat U_\sigma|$ and average over these samples to determine the response function estimate and covariance matrix.  In contrast to the SPA+RPA, the SMMC includes all fluctuations of the time-dependent auxiliary fields $\sigma$. 

We will discuss our SPA+RPA method and MEM approach in more detail in a forthcoming article~\cite{fanto_unpub}.

{\it Application to lanthanide nuclei} -- We calculated finite-temperature $M1$ $\gamma$SFs in a chain of even-mass samarium isotopes $^{148-154}$Sm with the SPA+RPA and used the MEM to refine these strength functions. We used the single-particle model space and Hamiltonian of Ref.~\cite{fanto2021} with a pairing plus quadrupole two-body interaction.  The $M1$ operator has the form
\be\label{M1}
\hat{\mathcal{O}}_{M1} = \sqrt{\frac{3}{4\pi}} \frac{\mu_N}{\hbar c} \left( g_l \mathbf{l} + g_s \mathbf{s}\right)\,,
\ee
where $\mathbf{l}$ and $\mathbf{s}$ are the orbital and spin angular momentum operators, respectively.  In our calculations, we used the free-nucleon $g$ factors $g_{l,p} = 1$, $g_{l,n} = 0$, $g_{s,p} = 5.5857$, and $g_{s,n} = -3.8263$.

\begin{figure*}[ht!]
\includegraphics[width=\textwidth]{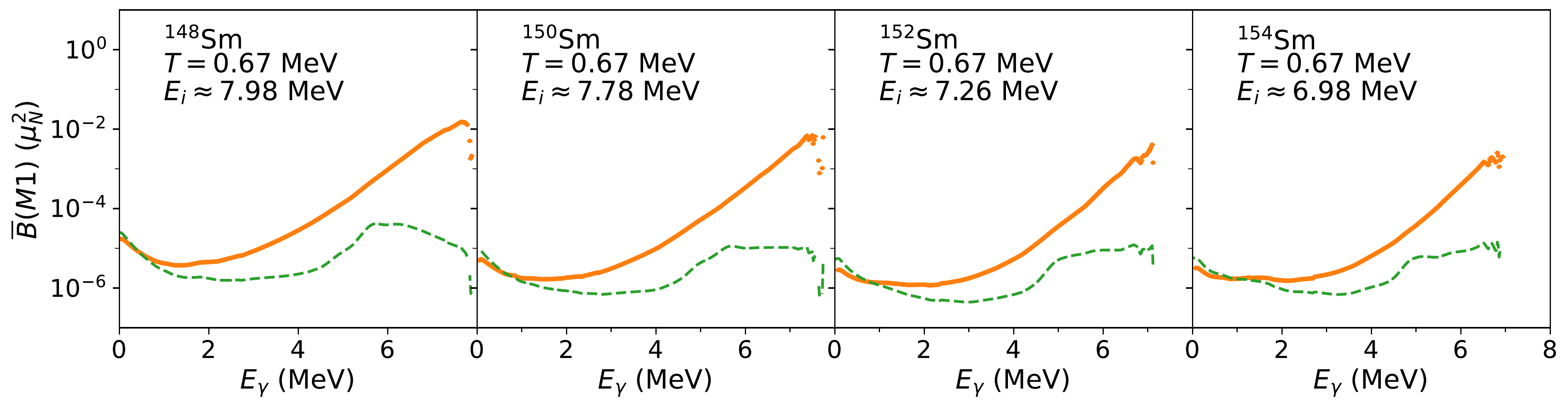}
\caption{\label{avgB-figure} The average $B(M1)$ values as a function of emitted $\gamma$-ray energy $E_\gamma$ for the even-mass samarium isotopes $^{148-154}$Sm at an average initial energy near their neutron separation energy.  The MEM results (green dashed lines) are compared with the SPA+RPA results (orange dots). We observe the LEE in all four isotopes. }
\end{figure*}

In Fig.~\ref{gsf-figure}, we show the finite-temperature $M1$ $\gamma$SF.  We compare the MEM $\gamma$SF results (green dashed lines) with the SPA+RPA results (orange dots) and SPA results (blue circles). The top row in Fig.~\ref{gsf-figure} shows temperatures at which the average initial energy in each isotope is near the neutron separation energy, while the bottom row shows low temperatures at which the isotopes are essentially in their ground states.  Positive values of $\omega$ correspond to the absorption of $\gamma$-rays.  

The reliability of the MEM depends on a good choice for the prior strength function.  In Fig.~1 of the Supplemental Material~\cite{supp} we show the $M1$ imaginary-time response functions  for the SPA and the SPA+RPA in comparison with the exact SMMC response function (which essentially coincides with the response function of the MEM strength function). We find that the SPA+RPA response is closer to the SMMC response than the SPA response and is quite close to the SMMC response. This indicates that our MEM results are reliable. 

In Fig.~\ref{gsf-figure}, we replaced the $\delta$ functions in the $\sigma$-dependent strength functions (\ref{sigma_gsf}) with Lorentzians of fixed width $
\epsilon = 0.2$ MeV.  This width is comparable to the bin width in final energy used in previous CI shell-model studies~\cite{schwengner2013,karampagia2017,schwengner2017,sieja2017,mitdbo2018}.  

For transitions starting near the neutron separation energy, the main effect of the MEM is to increase the height of the peak at $\omega \approx 0$.  At higher $\omega$ values, the MEM is in excellent agreement with the SPA+RPA result, which was used as a prior.  As the neutron number increases, the strength of this $\omega \approx 0$ peak is reduced, and a small excitation around $\omega \approx 2$ MeV emerges in the $\gamma$SF.  The location of this peak corresponds roughly to the location of the scissors mode observed in transitions from the ground state~\cite{bohle1984,heyde2010}, and we therefore tentatively interpret it as the scissors mode built on excited states~\cite{krticka2011}.  We also note an excitation at $\omega \approx 6$ MeV with a strength that is roughly independent of neutron number.

In the $\gamma$SF from the ground state shown in the bottom row of Fig.~\ref{gsf-figure}, we observe no zero-$\omega$ peak. Instead, as neutron number increases,  strength transfers from an excitation at $\omega \approx 6$ MeV, which we interpret to be the spin-flip mode, to a multi-humped low-energy excitation at $\omega \approx 1-3$ MeV.  This is roughly consistent with the appearance of the scissors mode with increasing deformation in heavy even-mass nuclei~\cite{heyde2010}. 

To investigate the LEE, we calculate the average value of $B(M1)$ as a function of the $\gamma$-ray energy from the finite-temperature $\gamma$SF.  We estimate the average $B(M1)$ value for an initial excitation energy $E_i$ and an emitted $\gamma$-ray energy $E_\gamma$ using
\be\label{avg_bvalue}
\overline{B}(M1; E_i, E_\gamma) \approx \frac{S_{M1}(T; \omega=-E_\gamma)}{\tilde \rho(E_i -E_\gamma)}\,,
\ee
where $T$ is the temperature describing an average excitation energy of $E_i$ and $\tilde \rho(E_x)$ is the total level density.  The latter is calculated from the total SMMC state density plus the spin-cutoff model.  As the LEE is a feature of compound-nucleus decay, we focus on downward transitions.  In Fig.~\ref{avgB-figure}, we show the average $B(M1)$ values versus  the emitted $\gamma$-ray energy $E_\gamma$ for average initial energies $E_i$ near the neutron separation energy.  In the MEM results (green dashed lines), we observe a LEE structure below $E_\gamma \approx 2$ MeV in each of the isotopes. This LEE is also present in the SPA+RPA results (orange dots) but is not as pronounced. We note that the  response function is not sensitive to the strength function at high values of $\omega$ because of the exponential suppression in the Laplace transform (\ref{response_integral}). On the other hand, the low $\omega$ results for the MEM strength function (for which the zero-$\omega$ peak is observed) are expected to be reliable and thus our MEM description of the LEE is likely to be accurate. 

\begin{figure}[]
\includegraphics[width=0.4\textwidth]{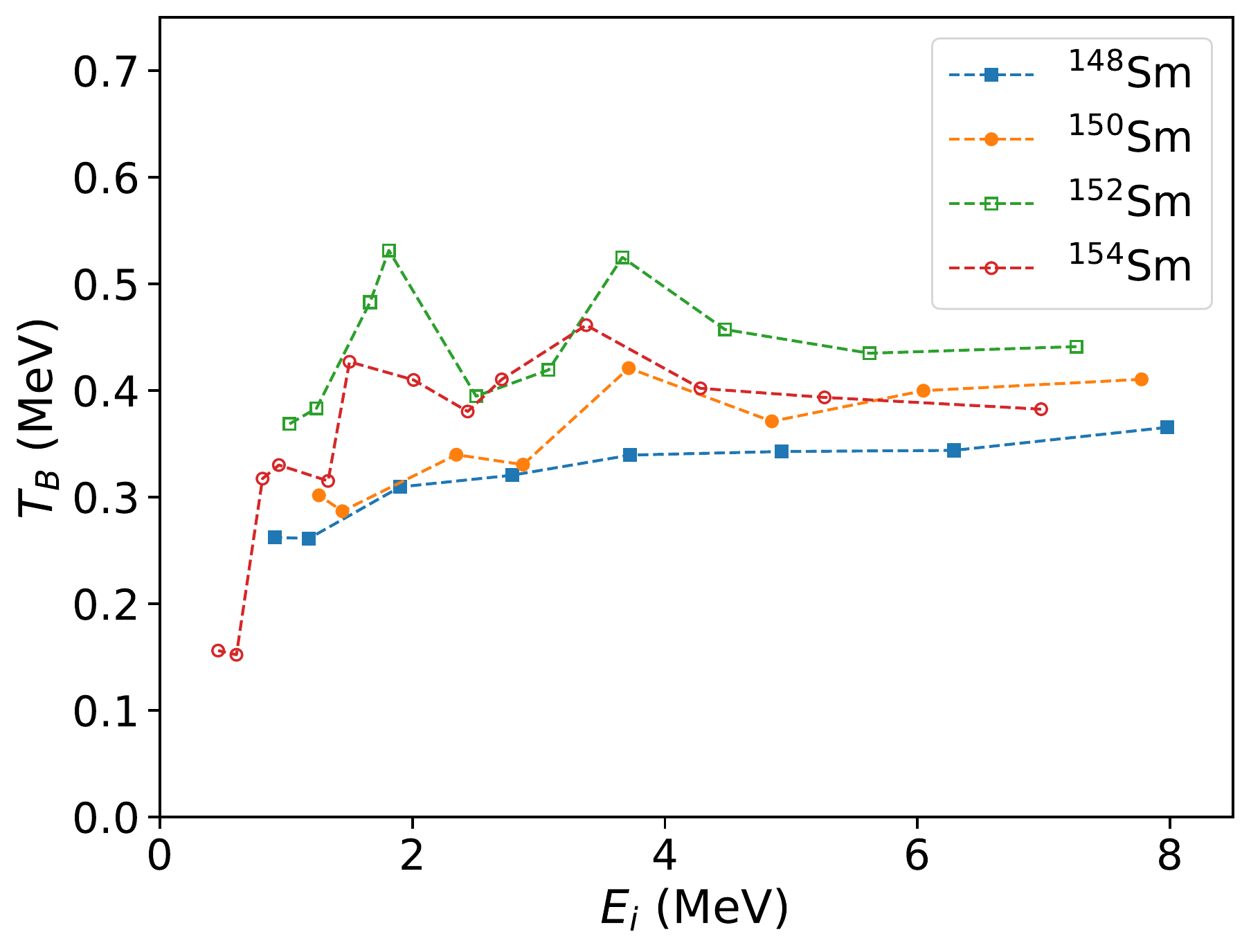}
\caption{\label{TB-figure} The $T_B$ parameter of the exponential form $B_0 e^{-E_\gamma/T_B}$ fit to the average $B(M1)$ results from the MEM at low $E_\gamma$ for various average initial energies $E_i$.  Results are shown for the even-mass samarium isotopes $^{148-154}$Sm.}
\end{figure}

Following Refs.~\cite{schwengner2013, karampagia2017}, we fit an exponential form $B_0 e^{-E_\gamma/T_B}$ to the average $B(M1)$ values at low $E_\gamma$ for various temperatures, each of which corresponds to an average initial energy $E_i$ for the transition.  We find that this exponential form provides a good description of the LEE structure.  In Fig.~\ref{TB-figure}, we show the fitted values of $T_B$ for $^{148-154}$Sm as a function of $E_i$.  We find that the $T_B$ values in each isotope remain roughly constant over a wide range of $E_i$ values.  This independence of the slope of the LEE as a function of the initial energy is consistent with conventional CI shell-model results in smaller model spaces~\cite{karampagia2017}.  The $T_B$ values for the isotopes are similar on average but are somewhat larger in the more deformed isotopes $^{152,154}$Sm, indicating that the LEE has a gentler slope in these isotopes.
This is consistent with CI shell-model results in open-shell iron nuclei~\cite{schwengner2017}.

{\it Conclusions} -- In this work, we applied the SPA+RPA to calculate finite-temperature $M1$ $\gamma$SFs in a chain of even-mass samarium isotopes $^{148-154}$Sm within the CI shell-model framework.  We also applied the MEM to derive the $M1$ $\gamma$SFs from the exact SMMC imaginary-time response functions using our SPA+RPA results as the prior strength. We calculated the average $B(M1)$ values as a function of the $\gamma$-ray decay energy for the even-mass samarium isotopes $^{148-154}$Sm and found a LEE structure. To our knowledge, this is the first theoretical description of a LEE in the $M1$ $\gamma$SF of heavy nuclei.  We showed that the LEE is roughly independent of the average initial energy, in agreement with previous results in medium-mass nuclei. If a LEE also exists in heavy nuclei near the neutron drip line, it is expected to enhance the radiative neutron capture cross sections, and thus alter considerably $r$-process nucleosynthesis.  We also observed the emergence of a structure consistent with the scissors mode both near the neutron separation energy and near the ground state as the neutron number increases.  

Finally, the methods we applied here are not specific to atomic nuclei, but are generally applicable to the calculation of strength functions in strongly interacting many-body quantum systems.

{\it Acknowledgments} --  This work was supported in part by the U.S. DOE grant No.~DE-SC0019521, 
and by the U.S. DOE NNSA Stewardship Science Graduate Fellowship under cooperative agreement No.~NA-0003960.
The calculations used resources of the National Energy Research Scientific Computing Center (NERSC), a U.S. Department of Energy Office of Science User Facility operated under Contract No.~DE-AC02-05CH11231.  We thank the Yale Center for Research Computing for guidance and use of the research computing infrastructure.


\setcounter{figure}{0} 

\onecolumngrid

\vspace*{1 cm}

\section{Supplemental material: Low-energy enhancement in the magnetic dipole $\gamma$-ray strength functions of heavy nuclei}

We applied the maximum-entropy method (MEM) to extract the finite-temperature $\gamma$-ray strength function ($\gamma$SF) from the imaginary-time response function calculated with the shell-model Monte Carlo (SMMC) method, using the static-path plus random-phase approximation (SPA+RPA) result as a prior.  As discussed in the main text and in Ref.~\cite{Sjarrell1996}, the MEM objective function combines a $\chi^2$ fit to the SMMC response function with an entropy term that biases the result toward the prior.  The method works best if the prior strength function provides a good initial approximation to the exact strength function.

\begin{figure*}[ht!]
\includegraphics[width=\textwidth]{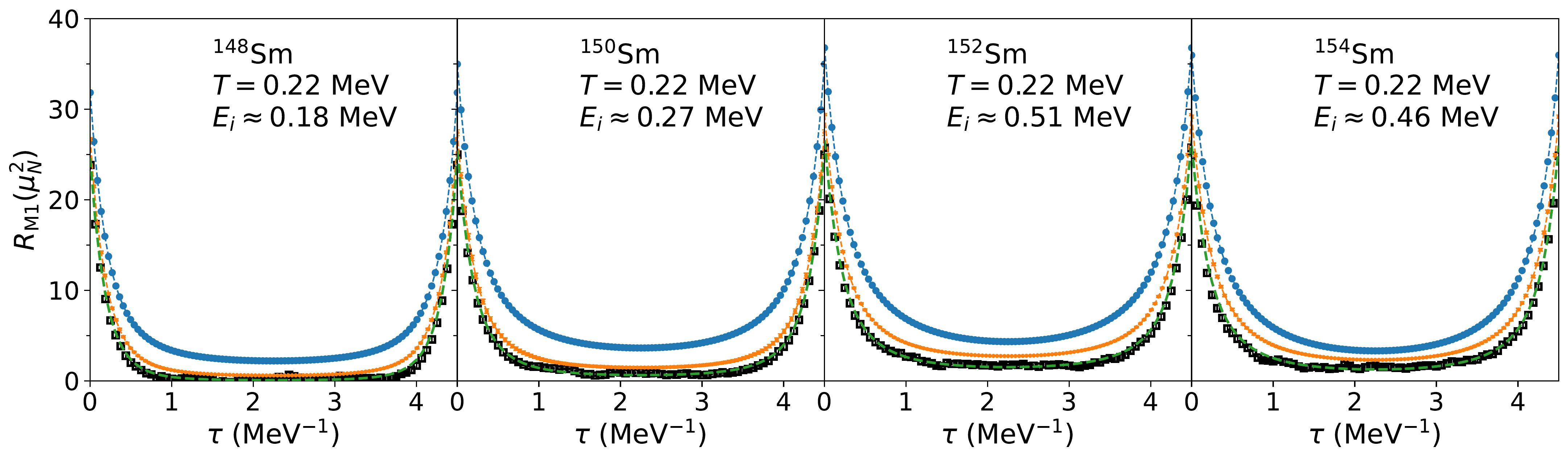}
\caption{\label{response-figure} The $M1$ imaginary-time response function $R_{M1}(\tau)$ as a function of imaginary time $\tau$ for a chain of even-mass samarium isotopes $^{148-154}$Sm.  The SPA+RPA response function (orange dots) is compared with the SMMC results (black squares).  The SPA (blue circles) and MEM (green dashed lines) result are also shown.}
\end{figure*}

To assess the initial agreement between the SPA+RPA and SMMC, we calculated the $M1$ imaginary-time response function $R_{M1}(\tau)$ in the SPA+RPA and compared it with the exact (up to statistical errors) $M1$ imaginary-time response function calculated in SMMC.  In Fig.~\ref{response-figure}, we compare the SPA+RPA response function (orange dots) with the SMMC response function (black squares) for the even-mass samarium isotopes $^{148-154}$Sm at a low temperature at which the nuclei are close to their ground states.  We include the SPA response function (blue circles) and the MEM fit result (green dashed lines) as well.  We find that the SPA+RPA provides better agreement than the SPA and is very close to the exact SMMC result.  The MEM is in nearly exact agreement with the SMMC response function, which is expected as the MEM fits the SMMC response.


\end{document}